# 13.16 Tbit/s Free-space Optical Transmission over 10.45 km for Geostationary Satellite Feeder-links

Annika Dochhan
*Advanced Technology*
ADVA Optical Networking SE
Meiningen, Germany
adochhan@advaoptical.com

Juraj Poliak
*Institute of Communication and Navigation*
German Aerospace Center (DLR)
Wessling, Germany
Juraj.Poliak@dlr.de

Janis Surof
*Institute of Communication and Navigation*
German Aerospace Center (DLR)
Wessling, Germany
Janis.Surof@dlr.de

Mathias Richerzhagen
*Institute of Communication and Navigation*
German Aerospace Center (DLR)
Wessling, Germany
Mathias.Richerzhagen@dlr.de

Helawae Friew Kelemu
*Institute of Communication and Navigation*
German Aerospace Center (DLR)
Wessling, Germany
Helawae.Kelemu@dlr.de

Ramon Mata Calvo
*Institute of Communication and Navigation*
German Aerospace Center (DLR)
Wessling, Germany
Ramon.MataCalvo@dlr.de

*Abstract*—We report a 13.16 Tbit/s 54-channel DWDM free-space optical data transmission field trial emulating the worst-case scenario for optical GEO satellite uplinks through the atmosphere, and show the system performance under turbulent conditions.

*Keywords—optical satellite communication, optical free-space communication, modulation*

## I. INTRODUCTION

Although data rates of terrestrial fiber networks and mobile communication base stations are increasing continuously, there are still unconnected areas, even in Europe and the US, and especially in South America and Africa. Digital connectivity significantly improves the standard of living by providing information, e.g. for health care applications, or by offering new business opportunities. Therefore, new solutions for providing data connection to each inhabited spot of the earth, regardless of its population density or geographical situation have to be found. Satellite communication from geostationary satellites (GEOs) can provide this connectivity [1]. However, the radio frequency (RF) bandwidth of the feeder-link between the gateway node on the ground and the satellites is limited to some GHz. Optical links offer several THz of bandwidth and no restrictions apply on specific frequencies, as it is the case for RF. Therefore, a scenario, where the feeder-link to the GEO is based on optical communication and the user-link providing internet connection to the users is based on RF, is a viable solution, especially in rural areas with low population density. To test such a connection, it is beneficial to employ a terrestrial testbed before establishing a real satellite feeder-link to investigate all influences due to atmospheric turbulences, which are especially challenging in the uplink to the satellite.

In this work a 10.45-km field testbed between the DLR site in Weilheim (Germany) and a site of the Germany's National Meteorological Service (DWD) at Hohenpeißenberg was used, to demonstrate the potential of optical communications for satellite feeder-links. Since this link is near to the ground and stays inside the atmosphere, the turbulence corresponds to the worst case for the uplink of an optical feeder link for geostationary satellites, which is much longer than 10.45 km. We transmitted 13.16 Tbit/s on 54 dense wavelength division multiplexed (DWDM) 50-GHz-spaced channels in the C-band. Besides the GEO feeder link scenario, the results can also be exploited for terrestrial free-space optical transmission, which might be an option to bridge rough terrain [2].

## II. SYSTEM SETUP

In the system setup, shown in Fig. 1, two channels under test (CUT) were considered. The signals for these channels are generated by commercially available coherent transponders (ADVA FSP 3000 Cloudconnect QuadFlex™), capable of modulating signals with DP-16QAM, DP-8QAM and DP-QPSK modulation formats. For the CUTs, a soft decision FEC with 25% overhead was included, leading to a gross data rate of approximately 275.8 Gbit/s, when 16QAM was used. The net data rate in this case is 200 Gbit/s and is halved for DP-QPSK, while the symbol rate is around 34 GBd. To generate the neighboring DWDM signals as loading channels, two optical modulator assemblies were employed, consisting of DP IQ modulators, drivers and generators for test sequences. Both assemblies operated at 245 Gbit/s DP-16QAM with 15% FEC overhead. To generate uncorrelated data on direct neighbors, each modulator assembly was fed by 26 or 25 of 100-GHz-spaced continuous wave lasers with a 50-GHz shift in between the two combs. The complete DWDM spectrum was generated by combining even and odd channels, the two CUTs and a third single-polarization (SP) 100-Gbit/s QPSK signal at 193.4 THz, the use of which will not be further discussed in this paper. Due to the high coupling losses when combining the laser and the insertion loss of the modulator assembly, the transmit optical signal-to-noise-ratio (OSNR) of the comb channels was below 12 dB. The three signals that were evaluated after transmission, namely the SP-QPSK signal and the CUTs, have significantly higher output power and hence transmit OSNR than each of the other channels when coupling them together. A programmable wavelength selectable switch (WSS) was used to equalize the channel power levels, leading to equal channel powers for all channels while maintaining the high OSNR for the SP-QPSK signal and the CUTs. The WSS is followed by two EDFAs which amplify the signal to a final transmit power of 32 dBm. The DWDM signal is launched into the optical ground station terminal and emitted into the air. A description of the transmission link of 10.45 km can be found in [1].

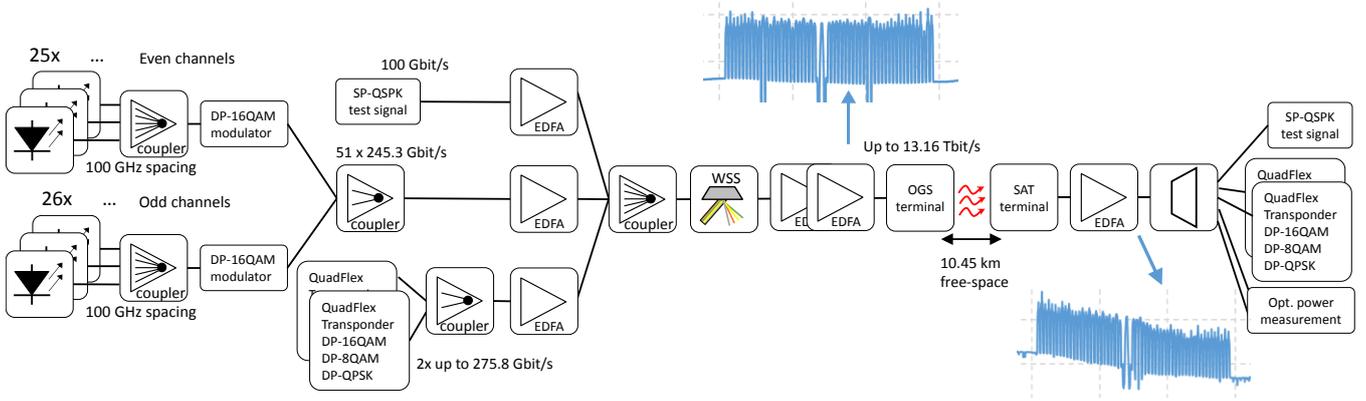

Fig. 1: Setup for optical free-space transmission of 54 DWDM channels, carrying up to 275.8 Gbit/s. Insets: TX and RX optical spectrum.

At the RX side, after the receiving telescope, the signal is coupled into standard single mode fiber using a fine-steering mirror controlled in closed loop by a four-quadrant-detector, as described in [1]. Back in the fiber, the signal is pre-amplified by an EDFA and de-multiplexed. The two CUTs are simultaneously captured and demodulated by the RX of the QuadFlexTM transponders. The SP-QPSK signal is received, and two further spectrally separated channels detect the received optical power to evaluate the atmospheric scintillation. The insets in Fig. 1 show the spectrum of the transmitted signal and the received spectrum at Hohenpeißenberg. The sweep time of the optical spectrum analyzer with ~1s is much slower than the turbulence induced variations of the received signal, thus an average of 50 sweeps is performed at the RX. The strong tilt of the spectrum can be related to two effects, the limited capability of the OSA to display the varying spectrum and the gain spectrum of the high power EDFA at the TX side.

### III. Measurement procedure and results

The quality of the transmitted signal was evaluated by means of bit error ratio (BER) measurements before forward error correction (pre-FEC BER). Unlike fiber transmission, the free-space optical link suffers severely from power fading due to the atmospheric turbulences occurring on a time scale of milliseconds. Although the average pre-FEC BER might be below the FEC limit over a longer period of time, data will be lost if the instantaneous BER raises significantly above the FEC limit during fading events. Therefore, it is necessary to continuously measure the BER over time with a time resolution smaller than the coherence time of the atmospheric turbulence. Given that turbulence effects are the same in all channels at a given time instance [1], it is possible to measure the channels under test sequentially.

The transponders used in this demonstration continuously measure the pre-FEC BER in intervals of 10 ms. The pre-FEC BER is calculated from the number of corrected errors by the FEC. An accurate value can only be obtained when no uncorrected FEC blocks occur and the post-FEC BER is zero. The pre-FEC BER limit for error-free detection was between 2e-2 and 3e-2. The transponders are controlled via a PC. Due to limited processing speed, the results of approximately sixteen 10 ms intervals can be read-out in 1 s, effectively leading to a time-sampling of the BER values. First, 16QAM is used on all channels. The CUTs are tuned through the whole spectrum, one on the even, one on the odd grid, replacing the neighboring channels at their frequency. The BER is measured over 2 min for each channel. Due to a relative long time required for necessary adjustments to the free-space optics and the receiver when changing wavelengths, one complete measurement cycle takes several hours. This leads to changing atmospheric conditions for each channel and has to be considered when comparing the results.

The measured BER values were processed as follows: In a first step, three types of 10-ms intervals are defined. If the post-FEC BER and the number of uncorrected blocks were both zero, the pre-FEC BER value is considered valid, and the transmission is error free. If the pre-FEC BER is "nan" (not a number), the data is lost, no matter how many uncorrected blocks are present. If the pre-FEC BER has any value and the number of uncorrected blocks is not zero, the calculated pre-FEC BER value is not valid and is discarded. However, it is very likely that only part of the 10-ms interval was affected by a fade. This happened often, especially during very strong scintillation, while the event of completely lost data was much rarer. To better capture these fades, 1-ms intervals would have been necessary, which were not possible due to processing time at the PC. In a real transmission, it is likely that most of the date with uncorrected blocks will be received correctly while only 1 ms of data might be lost. To account for this possibility, in the evaluation the data intervals with uncorrected blocks are displayed separately form the completely lost ones. To assess the quality of transmission, the percentage of each type of interval is evaluated and plotted in the upper part of Fig.2. The maximum probability of complete loss was 1.52% for the 194.75-THz channel. The lower part of Fig.2 shows the histograms of the pre-FEC BERs which are detected correctly to show the spread within a 2-min measurement window.

To verify and determine the influence of the spectral tilt at the RX, we estimated the mean OSNR for each channel. This is performed by averaging 50 spectral sweeps at the RX side during a time of very low scintillation. From the averaged spectrum, the OSNR of the center channel (the SP-QPSK signal) is determined. After finishing all free-space experiments, TX and RX were connected back-to-back, replacing the free-space link by an attenuator such that the OSNR of the center channel matched the value of the free-space link (20.89 dB). Then, the OSNR for each channel was measured.

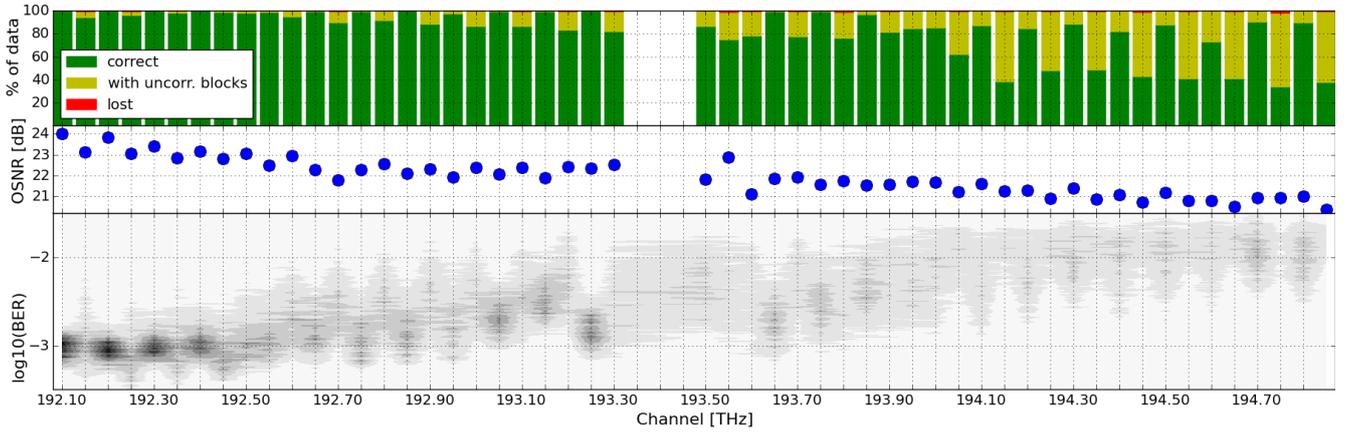

Fig. 2: Results of 13.16-Tbit/s data transmission over a 10.45-km free-space optical link for each DWDM channel. Upper part: Percentage of error-free data after FEC, data with uncorrected blocks after FEC, and completely lost data. Center part: Mean OSNR of each channel at the RX. Lower part: PDF of pre-FEC BER for correctly received data for each channel.

These results are shown in the middle of Fig. 2, partly explaining the low percentage of correctly received data at the higher frequency part of the spectrum. Regarding the strong difference between even and odd channels, it must be noted that neighboring channels have not been measured at the same time and thus have experienced different scintillations during the measurements. Finally, the modulation format for one CUTs was switched to DP-QPSK. Still, the two CUTs are captured simultaneously; therefore both channels experienced the same turbulence. Here, they are located at 193.6 THz and 193.55 THz, respectively. For 16QAM, 61.34% of the intervals were detected correctly, 38.61% showed uncorrected blocks and 0.05% were completely lost. For QPSK, the numbers are 99.54%, 0.41% and 0.05%, respectively.

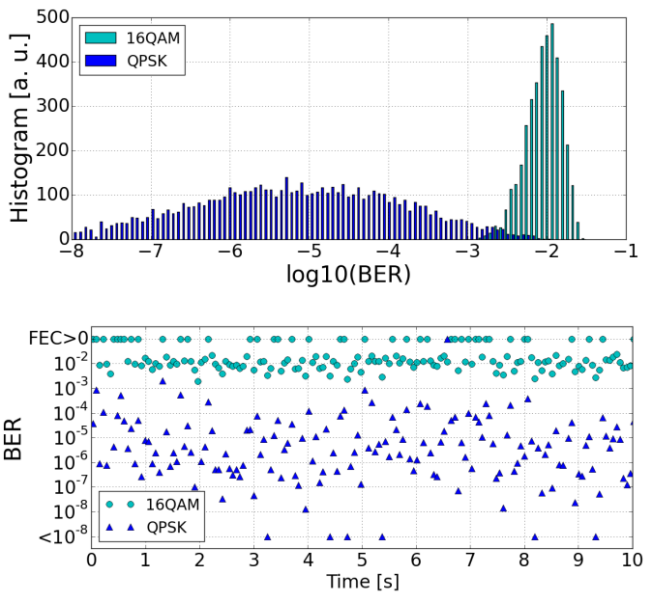

Fig. 3: Histogram (upper plot) and time evolution (lower plot) of pre-FEC BER for two simultaneously measured channels 16QAM and QPSK. FEC>0: Measurement with uncorrected blocks or no valid BER value.

Fig. 3 (top) shows the histograms of the pre-FEC BER of the corrected data intervals and the exemplary time evolution over 10 s for both formats (bottom). In case of QPSK, zero errors were detected in some 10 ms intervals and therefore the BER appears as zero (shown as <$10^{-8}$>), while 16QAM still encounters some errors. Regarding the time evolution, the general trend for both formats is the same: when QPSK exhibits a high BER, 16QAM cannot be detected anymore (FEC>0); when QPSK shows zero BER, 16QAM also shows a low BER value. It must be noted that the OSNR of the link is close to the required receive OSNR for 16QAM. Under static condition the required OSNR for 16QAM to be detected correctly was around 19 dB, leaving less than 2 dB margin for the worst channel according to Fig. 2.

IV. CONCLUSION

We demonstrated a 13.16 Tbit/s (51x245.3 Gbit/s + 100 Gbit/s +2x 275.8 Gbit/s) 10.45-km free-space optical (FSO) link using optical transponder designed for terrestrial WDM systems. The FSO scenario emulated a GEO satellite feeder-link but can also be used to reach rural areas or to complement terrestrial fiber-optic communication. The transmission performance in air is strongly impacted by atmospheric turbulence leading to an OSNR fluctuation. Quick adaptation to varying channel losses is key to maximize transmission capacity and minimize outage times.

ACKNOWLEDGMENT

We thank our former ADVA colleague Nicklas Eiselt for supporting us during preparation and during the measurement campaign. Also, we thank the staff of the German Weather Service (DWD) Hohenpeißenberg and the DLR Weilheim for their generous support and for sharing their premises during the outdoor measurements.